\documentclass{PoS}

\usepackage[center,footnotesize,hang]{subfigure}

\title{Lepton Flavour Violation Theory}

\ShortTitle{Lepton Flavour Violation Theory}

\author{\speaker{Thorsten Feldmann} \addtocounter{footnote}{1}\thanks{{Preprint IPPP/11/22, DCPT/11/44}} \\
IPPP, Department of Physics, University of Durham, Durham DH1 3LE, UK\\
E-mail: \email{thorsten.feldmann@durham.ac.uk}}

\abstract{I discuss theoretical and phenomenological aspects of flavour violation in charged lepton transitions.
This includes minimal extensions of the Standard Model within effective-theory approaches, 
as well as an incomplete selection of concrete new physics models.}

\FullConference{The 13th International Conference on B-Physics at Hadron Machines\\
                 April 4-8 2011\\
                 Amsterdam, The Netherlands}

\begin{document}

\section{Motivation}

\def\Br{{\cal B}}

The experimental observation of oscillations between different neutrino flavour eigenstates 
 implies that neutrinos are massive and that 
the lepton-family quantum numbers $(L_e,L_\mu,L_\tau)$
are not conserved. As a consequence, we also expect lepton-flavour violation (LFV) in
transitions between charged leptons. However, the physics responsible for 
neutrino masses and mixing -- in general -- can be independent of the physics
related to LFV in charged lepton processes:
\begin{itemize}
 \item Neutrino masses are most naturally obtained from see-saw scenarios, where the 
       masses of the new heavy particles are typically of the order of a
       grand-unification (GUT) scale, and the new interactions are generally also
       violating lepton-\emph{number} $L=L_e+L_\mu+L_\tau$ (LNV). The simplest candidates for such new particles
       are heavy right-handed Majorana neutrinos which naturally fit into 16-plets
        of an $SO(10)$ GUT. The dynamics of these heavy particles may also generate
       a baryon-antibaryon asymmetry in the universe (baryogenesis via leptogenesis).

  \item In a minimally extended Standard Model (SM), where the
     only source of LFV is coming from the operators responsible for the neutrino masses,
     the LF-violating effects are suppressed by loop factors and by the neutrino-mass differences, and turn out
     to be tiny. For instance, for $\mu\to e\gamma$ transitions, one obtains 
     $\Br[\mu \to e\gamma]_{\rm SM} \sim 10^{-54}$, to be compared with the present (or expected future)
     bounds $\Br[\mu \to e\gamma]_{\rm exp.} < 10^{-11(13)}$.

 \item On the other hand, generic models for new physics (NP) at the TeV scale
   contain new sources for LFV (but not necessarily for LNV), leading to decay rates accessible
   with future experiments \cite{Raidal:2008jk}.

\end{itemize}

From the low-energy point of view, these observations can be accounted for
by considering the SM as an effective theory and extending its Lagrangian,
\begin{equation}
 {\cal L}_{\rm eff} = {\cal L}_{\rm SM} + \frac{1}{\Lambda_{\rm LNV}} \, {\cal O}^{\rm dim-5} +
  \frac{1}{\Lambda_{\rm LFV}^2} \,  {\cal O}^{\rm dim-6}+ \ldots \,, 
\qquad \mbox{with} \ \Lambda_{\rm LNV} \gg \Lambda_{\rm LFV} \,.
\label{NP:gen}
\end{equation}
Here, the dimension-5 operator responsible for the neutrino masses is uniquely given 
in terms of the lepton doublets $L^i$ and the Higgs doublet $H$ in the SM,
\begin{equation}
   {\cal O}^{\rm dim-5} = (g_\nu)^{ij} \, 
     (\bar L^i \widetilde H)  (\widetilde H^\dagger L^j)^{c} + \mbox{h.c.} 
\end{equation}
and the misalignment between the flavour matrix $g_\nu$ and the Yukawa coupling
matrix $Y_E$ in the charged-lepton sector leads to a non-trivial mixing matrix 
$U_{\rm PMNS}$ for neutrino oscillations.\footnote{In a scenario with right-handed Majorana neutrinos
(type-I see saw), one would identify $g_\nu/\Lambda_{\rm LNV} = Y_\nu \, M^{-1} \, Y_\nu^T$,
where $Y_\nu$ is the Yukawa matrix in the neutrino sector, and $M$ the Majorana mass matrix.
Models with additional scalar triplets (type-II see saw) or fermion triplets (type-III see saw)
are also possible.}
An example for a dimension-6 operator, leading to LFV decays like $\mu \to e \gamma$, is
\begin{equation}
  {\cal O}^{\rm dim-6} \ni  c^{ij} \, \bar L^i \, \sigma^{\mu\nu} \, H  \, E_R^j \, F_{\mu\nu} \,,
\end{equation}
where $E_R$ are the charged-lepton singlets and $F_{\mu\nu}$ the hypercharge field strength tensor.
For generic coupling constants $c^{ij} \sim {\cal O}(1)$, the bound on the
$\mu\to e\gamma$ branching ratio would translate into $\Lambda_{\rm LFV} > 10^5$~TeV.

\begin{figure}[h]
\centering
\begin{tabular}{ccc} 
 \small \quad LLLL &  \small \quad  LLRR &  \small \ LR radiative \\[0.1em]
\includegraphics[width=0.305\textwidth]{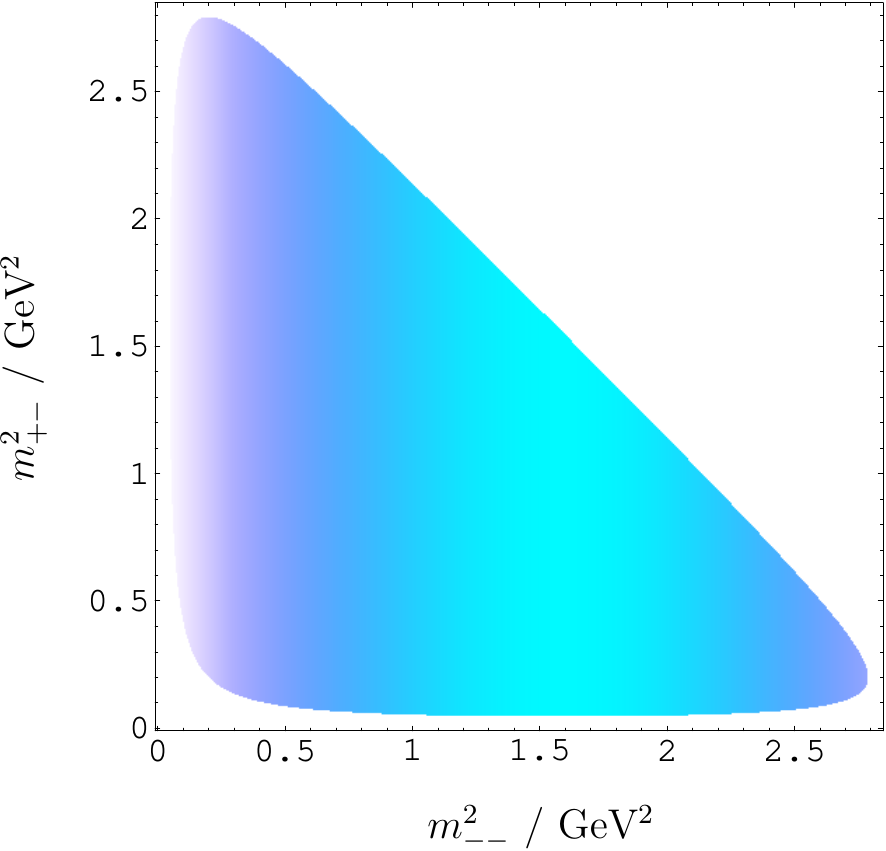}  &
\includegraphics[width=0.305\textwidth]{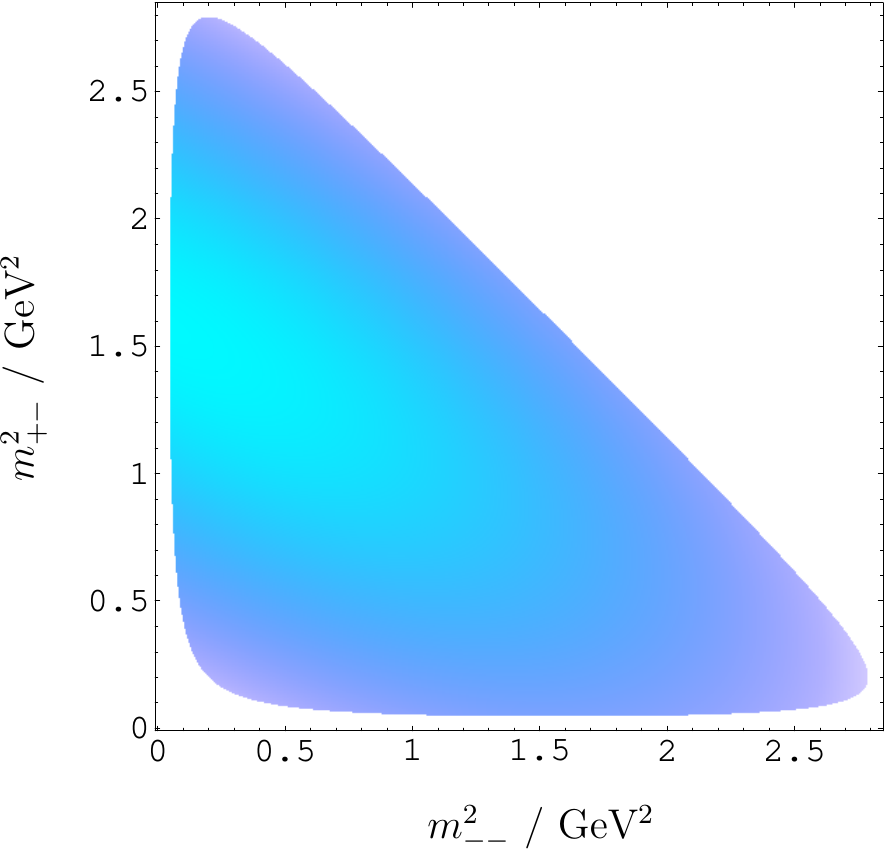}  &
\includegraphics[width=0.305\textwidth]{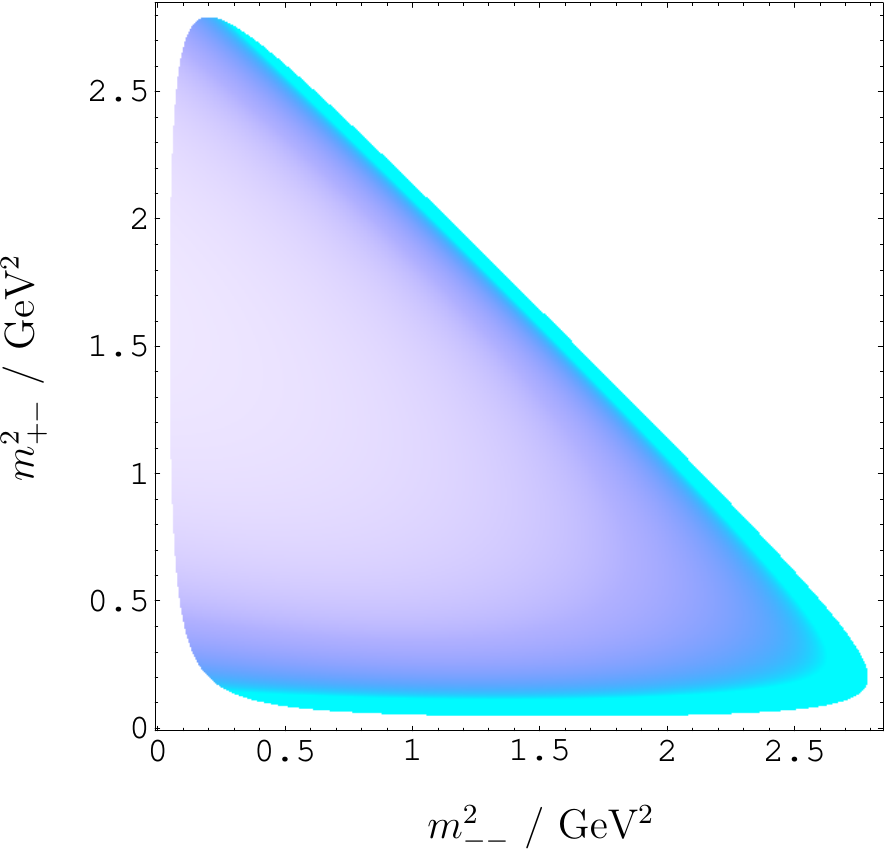} 
\end{tabular}
\caption{\label{fig:dalitz} Phase-space distributions for $\tau \to 3\mu$ from different chiralities in NP operators:
4-lepton operators with only left-handed leptons (LLLL), 4-lepton operators with 2 left- and 2 right-handed leptons (LLRR),
2-lepton operator (LR) with radiative decay $\tau \to \mu \gamma^* \to 3\mu$. Figure taken from \cite{Dassinger:2007ru}.}
\end{figure}

Besides the radiative decays $\mu \to e\gamma$ and $\tau \to \mu(e)\gamma$, the LFV
decays $\tau \to 3 \ell$ ($\ell =\mu,e$) and $\mu \to 3 e$ are important probes of
physics beyond the SM. Depending on which of the operators in (\ref{NP:gen}) dominates the decay, 
one obtains different distributions in invariant masses (see Fig.~\ref{fig:dalitz}) or
angular variables (see \cite{Giffels:2008ar}), a feature which also have to be accounted for when
deriving experimental limits. Finally, LFV can also be observed in hadronic decays and via
$\mu$-$e$ conversion in nuclei. For a summary of the experimental status and prospects, see
\cite{Lafferty:2011}.

\section{Minimal Lepton-Flavour Violation}

The idea of minimal flavour violation in the lepton sector (MLFV \cite{Cirigliano:2005ck}) is to expand the
flavour coefficients of dim$\geq 6$ NP operators in (\ref{NP:gen}) in terms of the flavour matrices
$Y_E$ and $g_\nu$ of the minimally extended SM, and to assume that the expansion coefficients are
at most of ${\cal O}(1)$. In this way, the flavour coefficients can be expressed in terms of PMNS-matrix
elements and lepton masses. Compared to the generic case, LFV processes are thus naturally suppressed.
On the other hand, with respect to the minimally extended SM, one gains factors of $\Lambda_{\rm LNV}/\Lambda_{\rm LFV}$
on the amplitude level.\footnote{The leading effects can be systematically singled out using a non-linear spurion
formalism \cite{Feldmann:2008av}.} For instance, the (dominant) coefficient of a purely left-handed 4-lepton operator, 
$$
  (\bar L_i \gamma^\mu  L^j) \; (\bar L_k\gamma_\mu L^l) \,,
$$
can be constructed from the flavour matrix $g_\nu$ which transforms as a 6-plet under the
$SU(3)_L$ flavour symmetry, and 
in MLFV it would be expressed as \cite{Cirigliano:2005ck,Dassinger:2007ru}
$$
 \left( c_8 \, \Delta^i_j \delta^{k}_l + c_{27}  \, G^{ik}_{jl} \right)\,,  \qquad c_{8,27} \sim {\cal O}(1) \,,
$$
where $\Delta$ and  $G$ denote the 8-tet and 27-plet 
in the reduction of  $g_\nu \otimes g_\nu^\dagger \sim \bar 6 \otimes 6 = 1 \oplus 8 \oplus 27$.

\begin{figure}[h]
\centering
\includegraphics[width=0.93\textwidth]{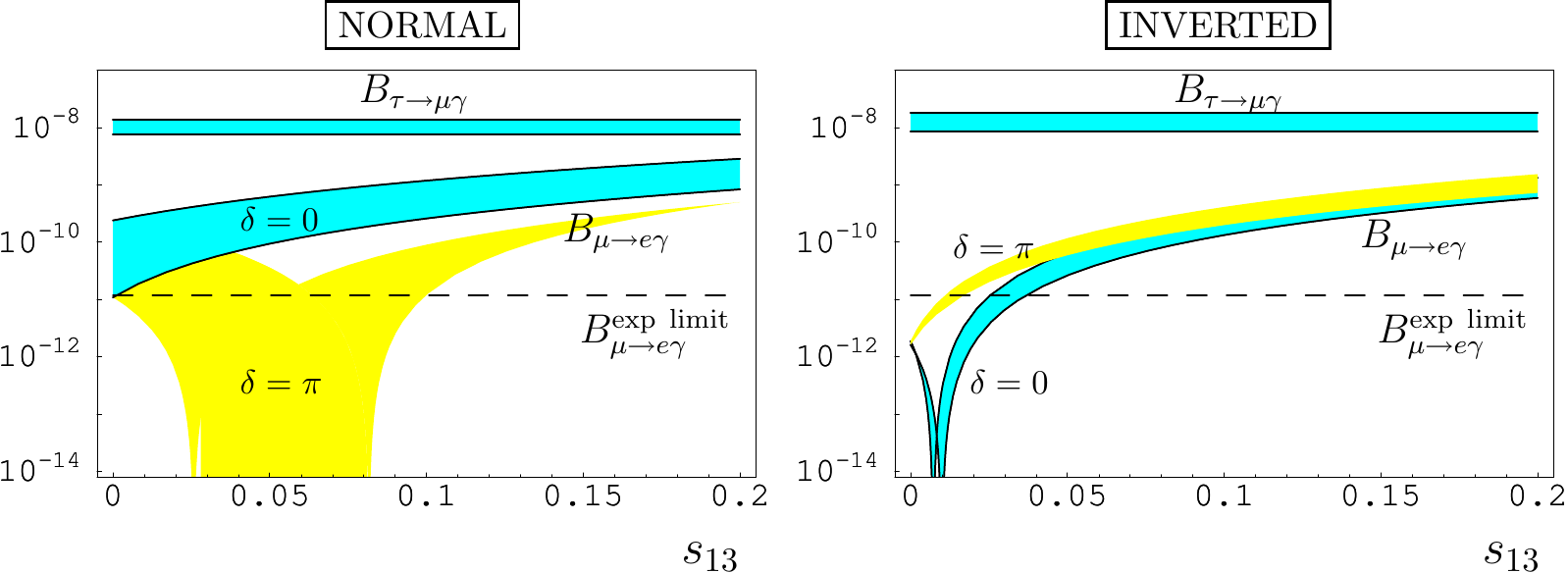}
\caption{\label{fig:MLFV} Branching ratios for $\mu \to e\gamma$ and $\tau \to \mu\gamma$ within MLFV compared to the experimental limits as a function
of the mixing angle $s_{13}=\sin\theta_{13}$ and the CP-phase $\delta$ in the PMNS matrix for normal or inverted neutrino-mass
hierarchy, assuming
$\Lambda_{\rm LNV} = 10^{10} \cdot \Lambda_{\rm LFV}$. Figure taken from \cite{Isidori:2009px}.}
\end{figure}

A typical prediction of MLFV is shown in Fig.~\ref{fig:MLFV}, where the branching ratios for
$\mu \to e\gamma$ and $\tau \to \mu\gamma$ are compared to the experimental limits as a function
of the neutrino mixing angle $\theta_{13}$. Notice that the scale $\Lambda_{\rm LNV}$ drops
out from the \emph{ratio} $\Br(\tau\to\mu\gamma)/\Br(\mu\to e\gamma)$, and therefore -- given the
existing/foreseen experimental bounds -- in MLFV one expects better experimental prospects to observe $\mu\to e\gamma$
than $\tau\to\mu\gamma$. It should be mentioned, however, that different mechanisms responsible for the generation
of neutrino masses can also lead to different formulations of MLFV with different phenomenological consequences,
see, for instance, \cite{alternatives}.

\section{LFV in Specific NP Models}

For a variety of NP models, the phenomenology of LFV observables has been
worked out in detail. For a comparison between different models, one is interested in
(i) whether the LFV operators are induced at tree-level or via loop processes, (ii)
how and to what extent the coefficients deviate from MLFV, (iii) how the predictions
compare with the present/foreseen experimental bounds, (iv) what the constraints are 
on new sources of LFV and new-particle masses, (v) how different LFV observables
are correlated. In the following, we discuss three  classes of NP models in somewhat
more detail: super-symmetric (SUSY) extensions of the SM, littlest Higgs models with T-parity, and a model with
a sequential 4$^{\rm th}$ generation. More information and references can be found
in \cite{Raidal:2008jk}.  
For further recent model analyses, see e.g.\ \cite{Akeroyd:2009nuetc}.

\subsection{LFV in SUSY Models}

In SUSY models (specifically the MSSM), new sources for LFV stem from
the soft SUSY-breaking sector, involving non-diagonal slepton mass matrices
and tri-linear couplings. The leading effects arise via sneutrino-chargino
and slepton-neutralino loops, with LFV triggered by the misalignment between
leptons and sleptons. Additionally, non-holomorphic couplings of the
Higgs doublets $H_{u,d}$ generate LFV coupling to neutral Higgs bosons which
become relevant if the ratio of vacuum expectation values $\tan\beta=v_u/v_d$
is large \cite{Paradisi:2005tk} (the same would be true for general 2-Higgs doublet models).
In the generic MSSM, it is useful to stick to the mass-insertion approximation,
assuming small off-diagonal entries in the slepton mass matrices. 
These can, for instance, be generated by considering specific SUSY-breaking scenarios 
with universal slepton parameters at high scales, and then working out the renormalization-group 
evolution to low energies. Depending on the model, additional 
constraints from (discrete) flavour symmetries \cite{Feruglio:2009hu} and/or specific assumptions 
on the see-saw parameters \cite{Antusch:2006vw}
can be implemented as well.

\begin{figure}[t!bhp]
 
\centering

\subfigure[$\tan\beta=2$, $u=0.01$ and $m_0=100$ GeV.]
   {\includegraphics[width=0.305\textwidth]{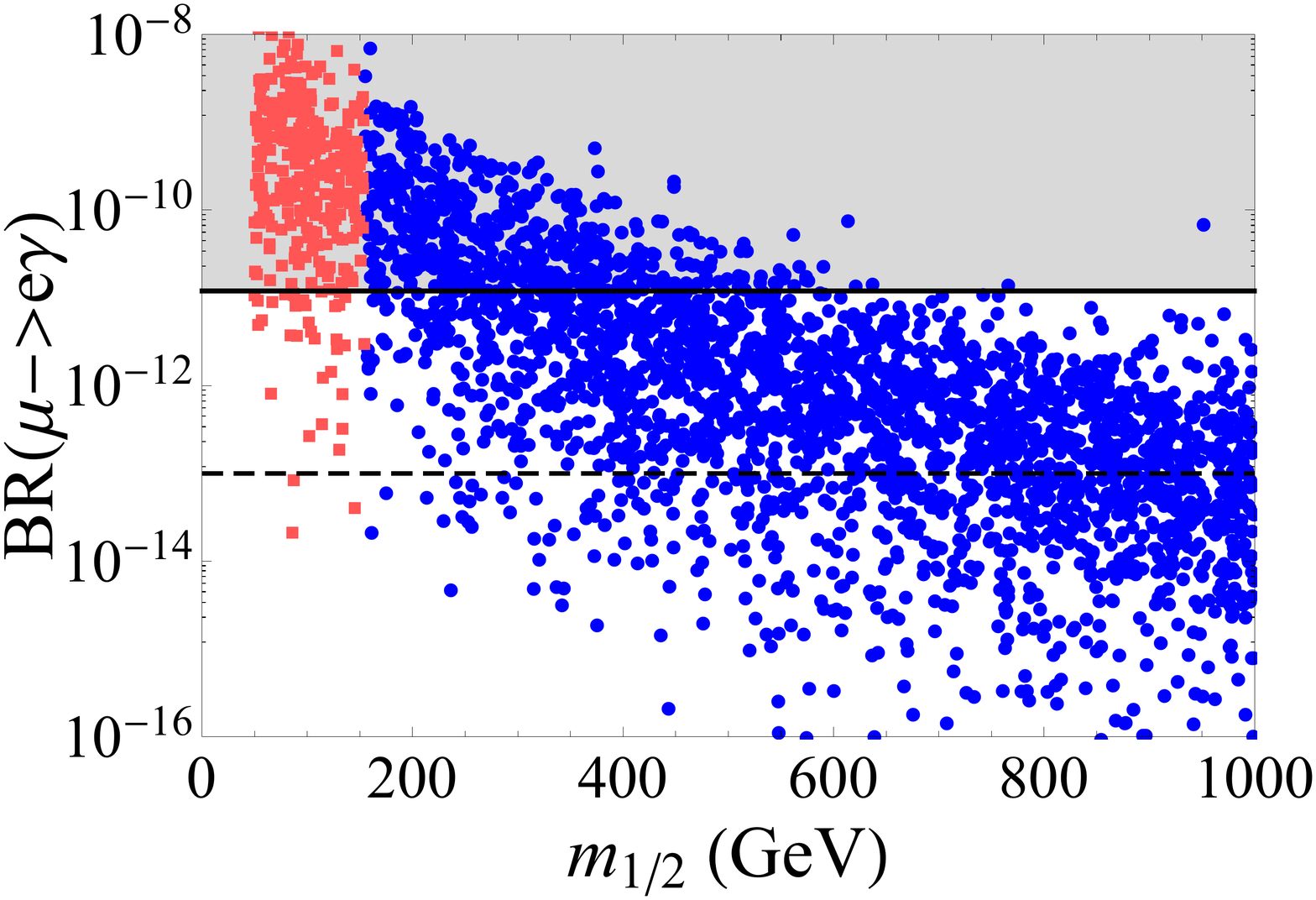}}
\subfigure[$\tan\beta=2$, $u=0.01$ and $m_0=1000$ GeV.]
   {\includegraphics[width=0.305\textwidth]{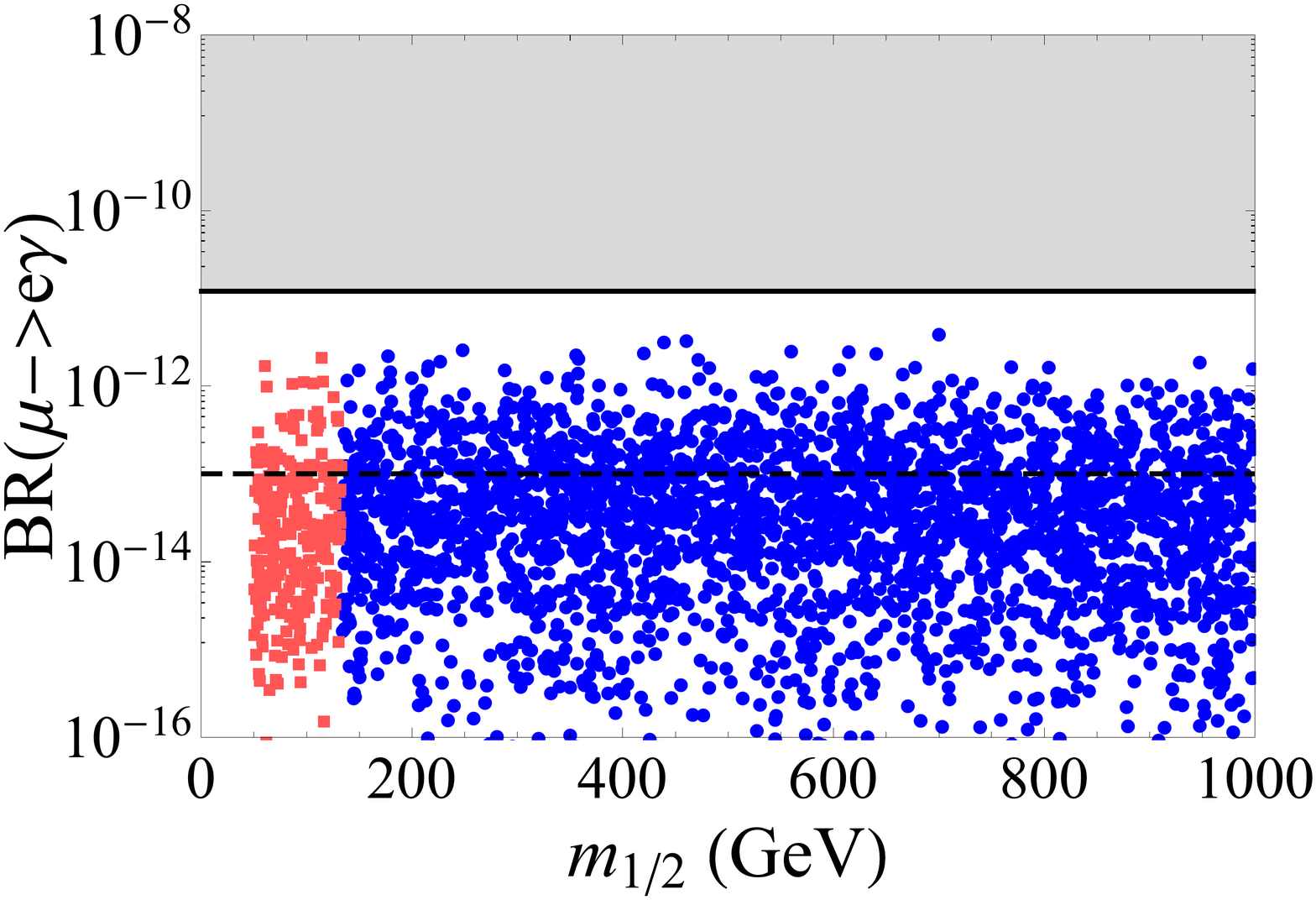}}
\subfigure[$\tan\beta=2$, $u=0.05$ and $m_0=100$ GeV.]
   {\includegraphics[width=0.305\textwidth]{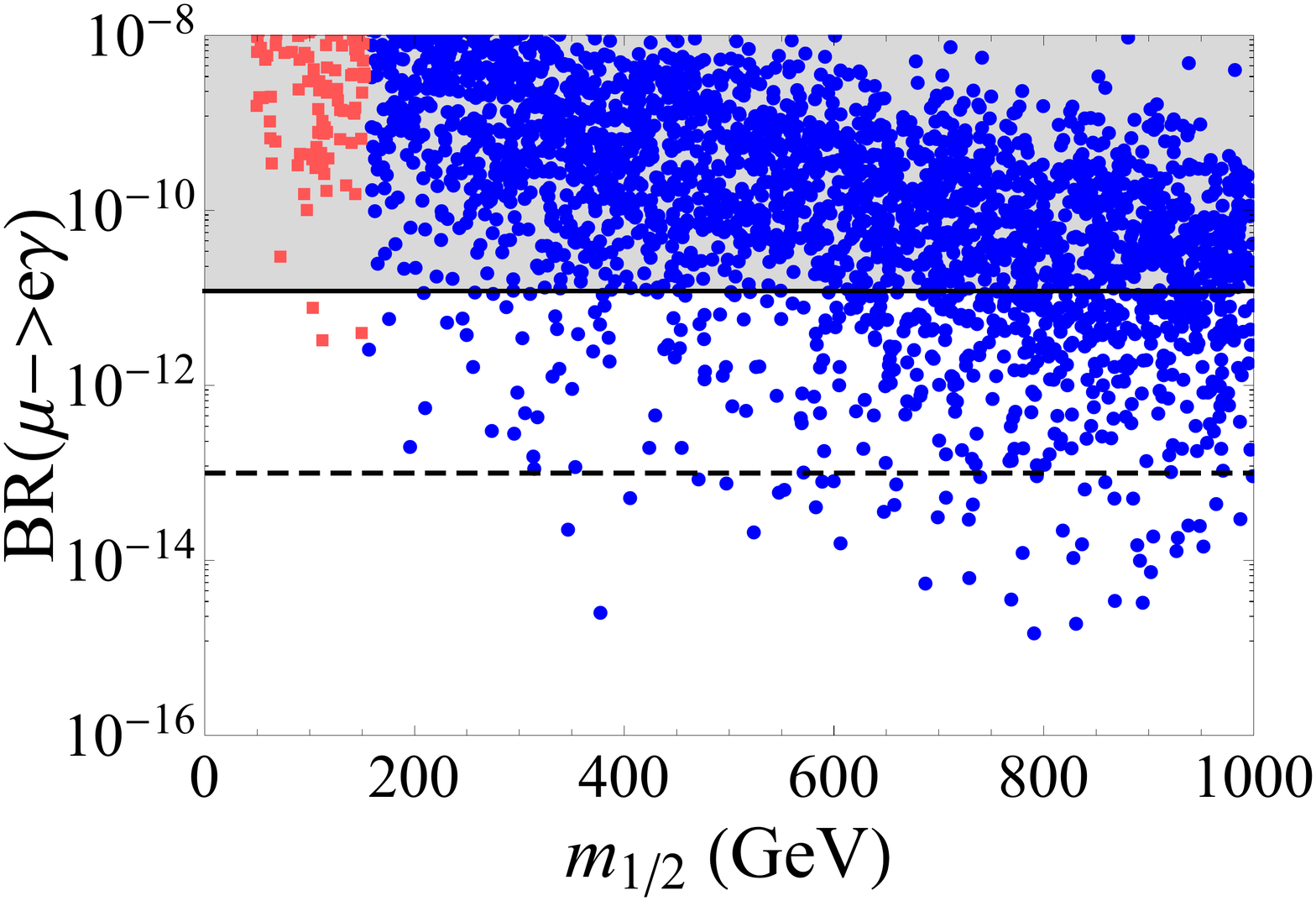}}
\subfigure[$\tan\beta=2$, $u=0.05$ and $m_0=1000$ GeV.]
   {\includegraphics[width=0.305\textwidth]{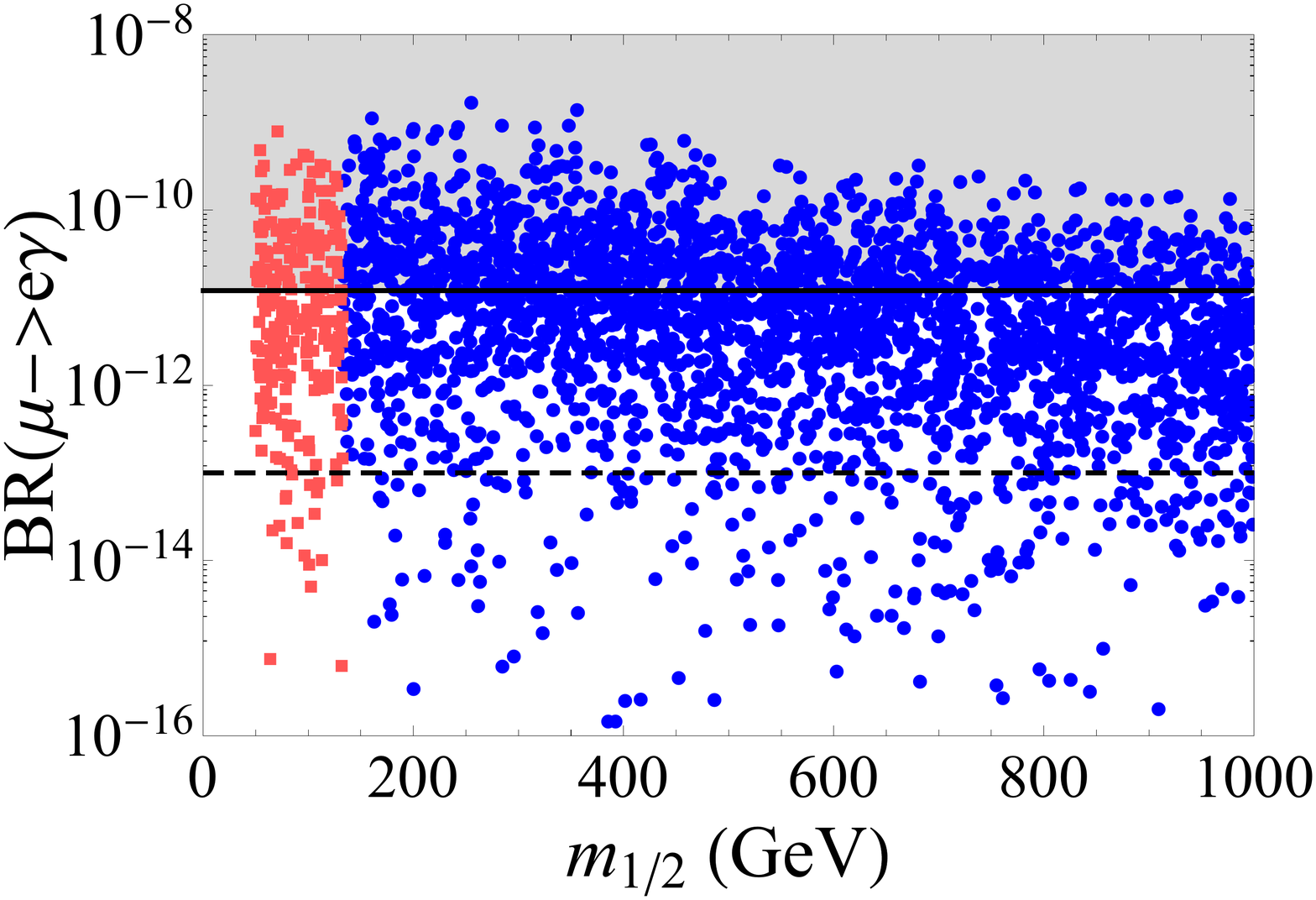}}
\subfigure[$\tan\beta=15$, $u=0.05$ and $m_0=100$ GeV.]
   {\includegraphics[width=0.305\textwidth]{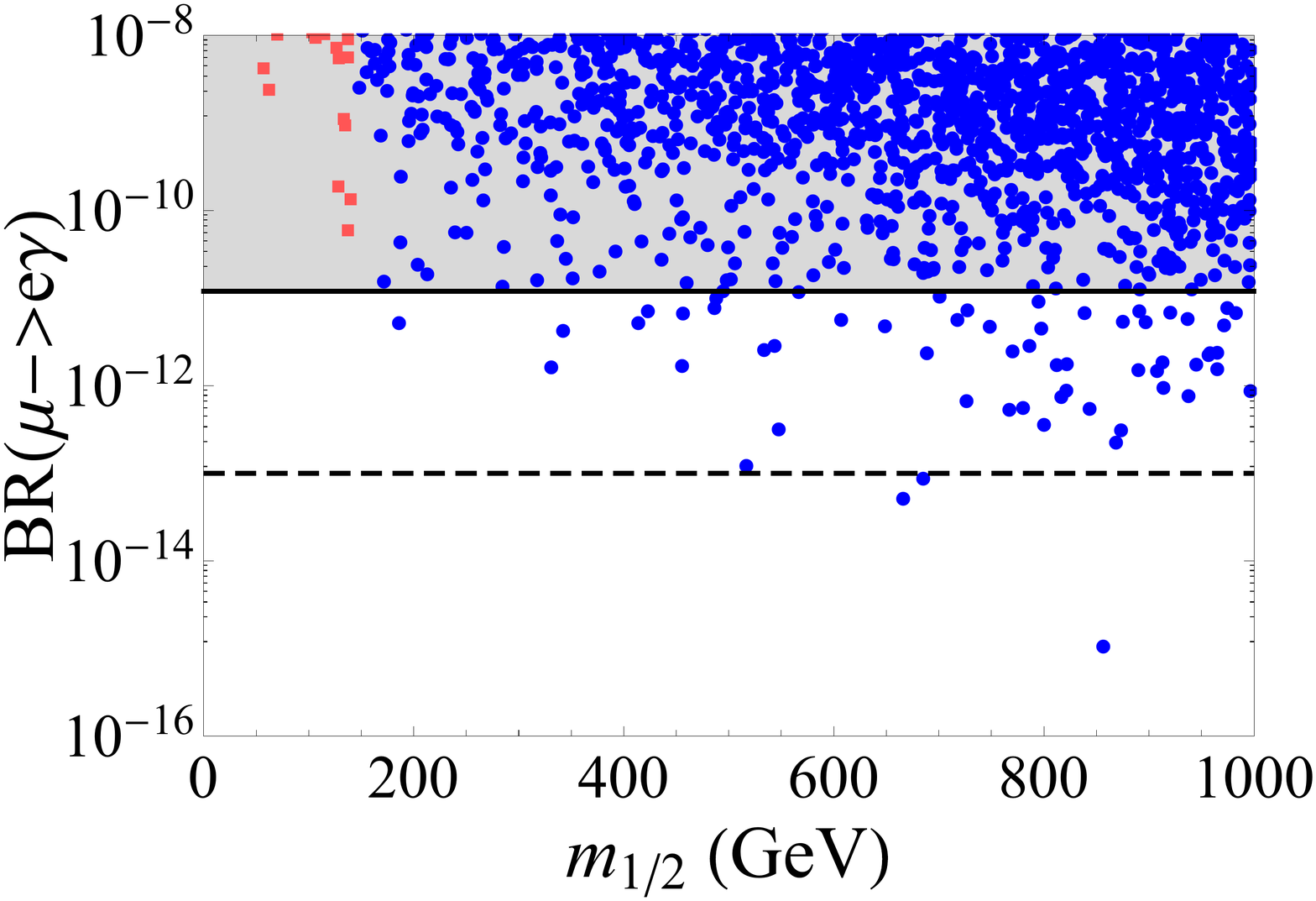}}
\subfigure[$\tan\beta=15$, $u=0.05$ and $m_0=1000$ GeV.]
   {\includegraphics[width=0.305\textwidth]{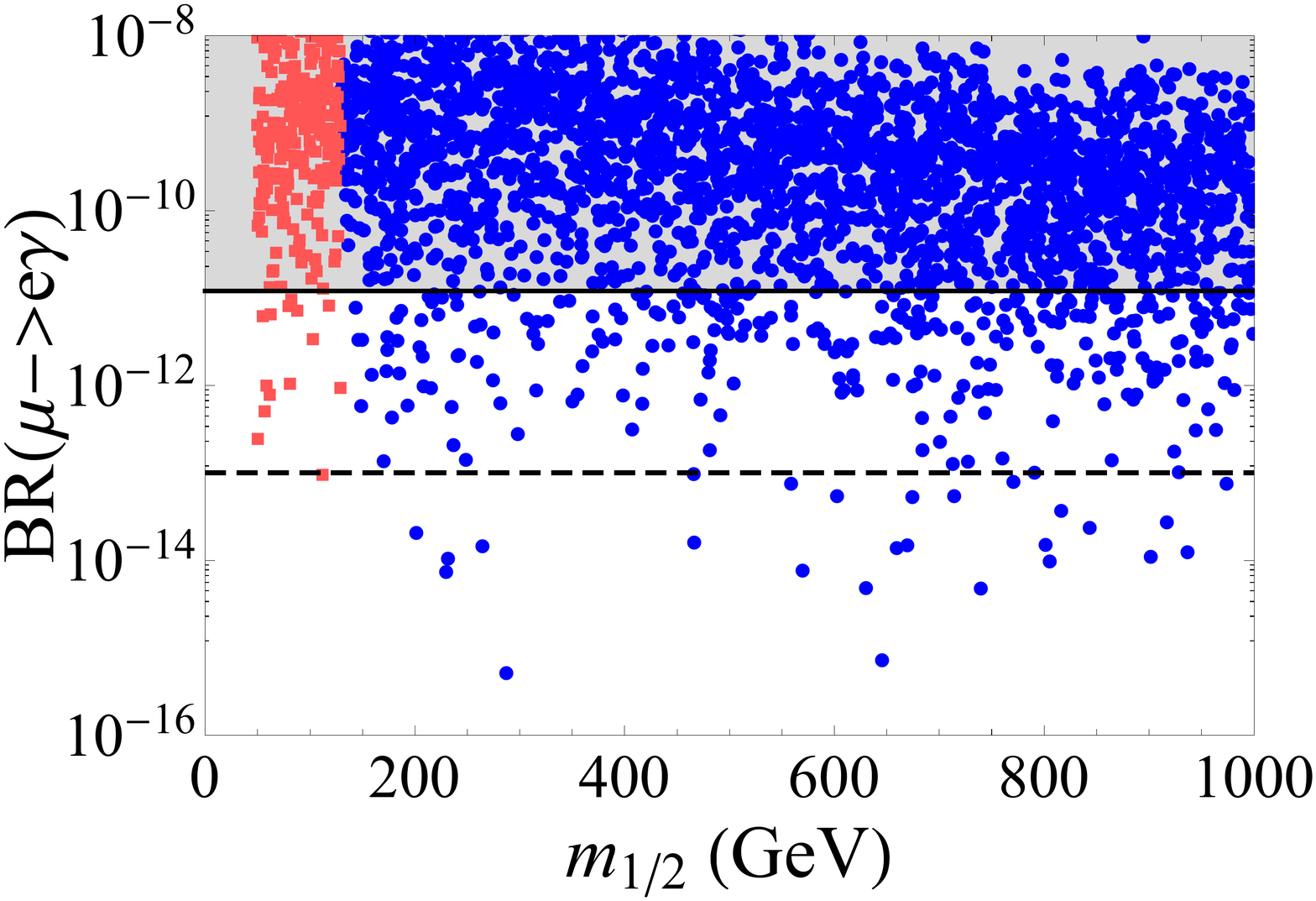}}
   
\caption{\label{fig:a4} Predictions $BR(\mu\to e \gamma)$ as a function of $m_{1/2}$, for different values of $\tan\beta$, $u$ and $m_0$,
in a SUSY model with discrete flavour symmetries. 
The red points correspond to the mass of the lightest chargino being below the limit coming from direct searches.
The horizontal lines show the current MEGA bound (continuous line) and the prospective MEG bound (dashed line).
Figures taken from \cite{Feruglio:2009hu}.}
\end{figure}

Let us, as an example, discuss a SUSY model in
\cite{Feruglio:2009hu} where the authors specify the symmetry group $A_4 \times Z_3 \times U(1)_{\rm FN}$,
       to enforce nearly tri-bi-maximal neutrino mixing.
Besides the SUSY mass parameters $m_{0,1/2}$ for sleptons and gauginos,
the relevant parameters for LFV phenomenology are
\begin{equation}
   u = \frac{\langle \phi_i \rangle}{\Lambda_f} \sim (0.01-0.05) \sim \theta_{13} \,,
\qquad t = \frac{\langle \theta_{\rm FN} \rangle}{\Lambda_f} \sim 0.05 \,,
\end{equation}
where $u$ is a small expansion parameter classifying the breaking of the $A_4$ symmetry,
and $\theta_{13}$ refers to the mixing angle in the PMNS matrix.
The parameter $t$ is responsible for the observed hierarchies in the charged-lepton Yukawa couplings,
with $\langle \theta_{\rm FN}\rangle$
breaking the Froggatt-Nielsen symmetry, and $\Lambda_f$ being a UV scale related to flavour-symmetry breaking.
In this scenario, the ratio of Higgs VEVs is restricted to small values, $2 \leq \tan\beta \leq 15$.
%
%
%
As a distinctive feature of the model, which allows a discrimination with respect to other SUSY constructions,
the flavour symmetries determine the structure of the slepton mass matrices, 
which contain off-diagonal entries at the flavour-symmetry breaking scale $\Lambda_f$. 
The ratios
$
 R_{ij} = \Br(\ell_i \to \ell_j\gamma)/\Br(\ell_i \to \ell_j \nu_i \bar \nu_j)
$ 
 are predicted to be approximately universal, $R_{\mu e} \approx R_{\tau\mu} \approx R_{\tau e}$.
Given the experimental limit on $R_{\mu e}$, the decay $\tau \to \mu\gamma$ would thus not be observable
  in the foreseeable future.
Moreover, the experimental constraints require at least one of the following conditions to be met:
a small flavour-symmetry breaking parameter $u \sim 0.01$, small $\tan\beta$,
and or large SUSY mass parameters (above 1~TeV). Predictions for the $\mu\to e\gamma$ branching
ratios are shown in Fig.~\ref{fig:a4}.

\subsection{Littlest Higgs Model with $T$-Parity}

\begin{figure}[t!bph]

\centering
 \includegraphics[width=0.47\textwidth]{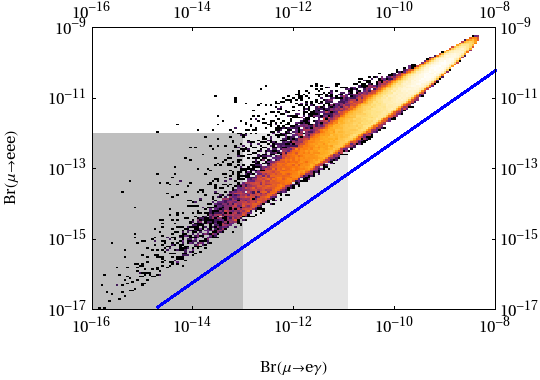} \qquad
 \includegraphics[width=0.47\textwidth]{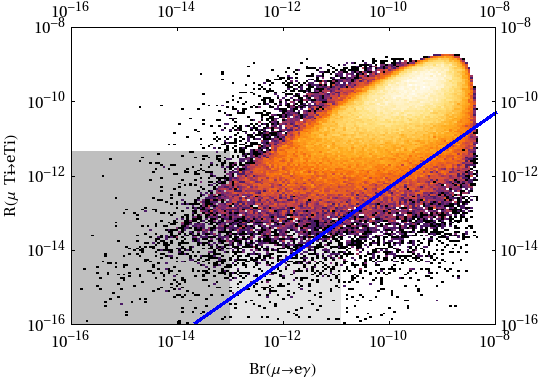}

\caption{\label{fig:lht} 
Correlations between $\mu \to 3e$ (left) or $\mu$-$e$ conversion (right)
 and $\mu \to e\gamma$ in the LHT model, for 
   $f=1~{\rm TeV}$, $300~{\rm GeV} \leq M_{H_i}^\ell \leq 1.5~{\rm TeV}$.
The blue dots denote the result one would obtain if only the dipole contribution
from $\mu \to e\gamma^*$ contributed. Figure taken from \cite{Blanke:2007db}.}

\end{figure}

An alternative to SUSY for resolving the hierarchy problem of the SM
is provided by Little Higgs models. In the following, we will refer
to an analysis of the Littlest Higgs Model with $T$ parity (LHT)
from Ref.~\cite{Blanke:2007db} (see also \cite{otherLHT}).
Besides new heavy gauge bosons (detectable at the LHC),
$T$-parity requires new heavy mirror leptons (and quarks) with masses of the
order TeV, which contribute to LFV processes via penguin and box diagrams.
The relevant input parameters of the LHT model are identified as
the scale parameter $f$, the three mirror lepton masses: $ M_{H_{1,2,3}}^\ell$,
the three mirror-lepton mixing angles: $\theta_{ij}^\ell$,
and three new (Dirac) CP phases $\delta_{ij}^\ell$.
In general, the potential LFV effects in the LHT model exceed the
SM by many orders of magnitude. Similarly as in the previous SUSY scenario, 
the present experimental constraints already require a certain amount of parameter tuning,
with a somewhat large LHT scale parameter,
and/or small mirror-lepton mixing angles,
and/or degenerate mirror lepton masses.
Examples for correlations between LFV decays and $\mu$-$e$ conversion
in the LHT model are shown in Fig.~\ref{fig:lht}.

\subsection{A $4^{\rm th}$ Generation of Leptons}

\begin{figure}[t!bph]

\centering
 \includegraphics[width=0.47\textwidth]{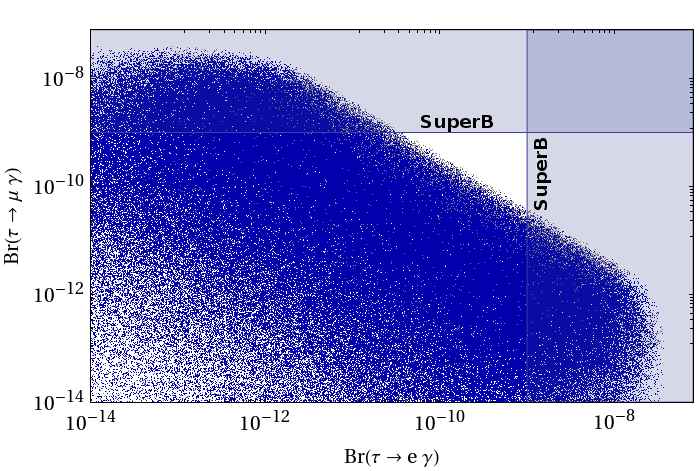} \qquad
  \includegraphics[width=0.47\textwidth]{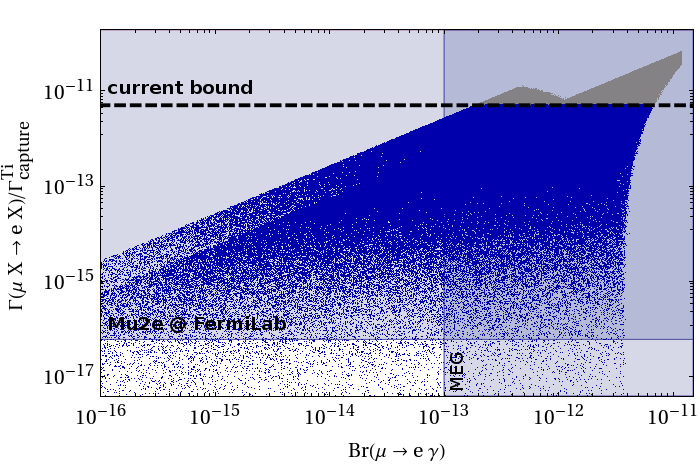}

\caption{\label{fig:4g}
Correlations between $\tau \to \mu\gamma$ and $\tau \to e\gamma$ (left) or $\mu$-$e$ conversion (right)
and $\mu \to e\gamma$ in the 4G model.
Figure taken from \cite{Buras:2010cp}.
}

\end{figure}

As a final example, we are going to discuss a model with an additional fourth generation (4G)
of leptons (and quarks), introducing a new heavy charged lepton $\tau'$ and a (Dirac-)neutrino $\nu_{\tau'}$,
together with an extended $4\times 4$ mixing matrix  $U_{ij}$ in the lepton sector
\cite{Lacker:2010zz,Buras:2010cp}.
In this set-up, the radiative $\mu$ and $\tau$ decays,  fulfill 
the simple relations
\begin{eqnarray}
\frac{\Br(\tau\to\mu\gamma)}{\Br(\mu\to e\gamma)}&\simeq&{\left|\frac{U_{\tau4}}{U_{e4}}\right|^2} \, 
\Br(\tau^-\to\nu_\tau\mu^-\bar\nu_\mu) \,, \cr 
\frac{\Br(\tau\to\mu\gamma)}{\Br(\tau\to e\gamma)}&\simeq &{\left|\frac{U_{\mu4}}{U_{e4}}\right|^2} \, 
\frac{\Br(\tau^-\to\nu_\tau\mu^-\bar\nu_\mu)}{\Br(\tau^-\to\nu_\tau e^-\bar\nu_e)}
\approx {\left|\frac{U_{\mu4}}{U_{e4}}\right|^2}\,, \cr 
\frac{\Br(\tau\to e\gamma)}{\Br(\mu\to e\gamma)}&\simeq &{\left|\frac{U_{\tau4}}{U_{\mu4}}\right|^2} \,
\Br(\tau^-\to\nu_\tau e^-\bar\nu_e) \,.
\end{eqnarray}
which put stringent constraints on the elements $|U_{i4}|$, independent of the heavy neutrino mass.
In turn, the rate for $\mu$--$e$ conversion in nuclei is directly proportional to $|U_{e4} U_{\mu4}|^2$.
This explains the correlations between radiative LFV decays and $\mu$-$e$ conversion
shown in Fig.~\ref{fig:4g}. From that figure, we conclude that both $\tau \to \mu\gamma$ and $\tau \to e\gamma$ may be
within experimental reach in the 4G model, but not simultaneously. Furthermore, concerning $\mu$-$e$ conversion
and $\mu \to e\gamma$, the foreseen experiments have the potential to further tighten the constraints on 
$|U_{e4} U_{\mu 4}|$.

A comparison for LFV branching ratios from various models
is shown in Table~\ref{tab}.

\begin{table}[h!tbp]
\caption{\label{tab} Comparison of ratios of LFV branching ratios in the LHT model \cite{Blanke:2007db},
the MSSM without significant Higgs contributions \cite{Ellis:2002fe},
the MSSM with significant Higgs contributions \cite{Paradisi:2005tk}, and
the 4G model \cite{Buras:2010cp}.
Table taken from \cite{Buras:2010cp}.
}
\begin{center} \def\text#1{{\rm #1}}
\renewcommand{\arraystretch}{1.5}
\small 
\begin{tabular}{|c|c|c|c|c|}
\hline
ratio & LHT  & MSSM (dipole) & MSSM (Higgs)& 4G \\\hline\hline
$\frac{\Br(\mu^-\to e^-e^+e^-)}{\Br(\mu\to e\gamma)}$  & \hspace{.8cm} 0.02\dots1\hspace{.8cm}  & $\sim6\cdot10^{-3}$ &$\sim6\cdot10^{-3}$ & $0.06\dots 2.2$ \\
$\frac{\Br(\tau^-\to e^-e^+e^-)}{\Br(\tau\to e\gamma)}$   & 0.04\dots0.4     &$\sim1\cdot10^{-2}$ & ${\sim1\cdot10^{-2}}$& $0.07\dots 2.2$\\
$\frac{\Br(\tau^-\to \mu^-\mu^+\mu^-)}{\Br(\tau\to \mu\gamma)}$  &0.04\dots0.4     &$\sim2\cdot10^{-3}$ & $0.06\dots0.1$& $0.06\dots 2.2$ \\\hline
$\frac{\Br(\tau^-\to e^-\mu^+\mu^-)}{\Br(\tau\to e\gamma)}$  & 0.04\dots0.3     &$\sim2\cdot10^{-3}$ & $0.02\dots0.04$& $0.03\dots 1.3$ \\
$\frac{\Br(\tau^-\to \mu^-e^+e^-)}{\Br(\tau\to \mu\gamma)}$  & 0.04\dots0.3    &$\sim1\cdot10^{-2}$ & ${\sim1\cdot10^{-2}}$& $0.04\dots 1.4$\\
$\frac{\Br(\tau^-\to e^-e^+e^-)}{\Br(\tau^-\to e^-\mu^+\mu^-)}$     & 0.8\dots2   &$\sim5$ & 0.3\dots0.5& $1.5\dots 2.3$\\
$\frac{\Br(\tau^-\to \mu^-\mu^+\mu^-)}{\Br(\tau^-\to \mu^-e^+e^-)}$   & 0.7\dots1.6    &$\sim0.2$ & 5\dots10& $1.4 \dots 1.7$ \\\hline
$\frac{{\rm R}(\mu\text{Ti}\to e\text{Ti})}{\Br(\mu\to e\gamma)}$  & $10^{-3}\dots 10^2$     & $\sim 5\cdot 10^{-3}$ & $0.08\dots0.15$&  $10^{-12}\dots 26$\\\hline
\end{tabular}
\end{center}
\end{table}


\section{Conclusions}

Lepton flavour violation in neutrino oscillations is a well-established phenomenon and naturally
related to lepton-number violating physics at the GUT scale, with neutrino masses 
suppressed by a large scale $\Lambda_{\rm LNV}$. As a consequence of the small neutrino masses,
LFV in charged lepton decays is tiny in the (minimally extended) SM. On the other hand, many
extensions of the SM with NP at the TeV scale predict sizable effects for lepton-flavour
violating muon or tau decays, as well as LFV hadronic decays or $\mu$-$e$ conversion in nuclei.
For the current and near-future experimental searches for LFV this implies that:
\begin{itemize}
 \item If LFV in charged lepton transitions is experimentally observed, this will be a clear
 signal for physics beyond the SM, and the correlations between different LFV observables can
 be used to distinguish different NP models.

 \item If, on the other hand, the experimental exclusion limits on LFV processes are further
    tightened, the parameter space of various NP models will become more and more constrained,
    and the absence of observable LFV beyond the SM would represent another case pointing
    towards a symmetry principle in the flavour sector being responsible for minimal flavour
    violation at low energies.
\end{itemize}

\acknowledgments

I would like to thank the organizers for a {\it Beauty}-ful conference, 
and Luca Merlo for useful comments on the manuscript.

\end{document}